\documentclass[12pt]{article}
\pdfoutput=1
\usepackage{putex}
\usepackage{feyn}

\usepackage{comment}
\usepackage{graphicx}
\usepackage{epstopdf}
\usepackage{enumerate}
\usepackage{cite}
\usepackage{tensor}
\usepackage{slashed}
\usepackage{feynmf}

\usepackage{hyperref}

\numberwithin{equation}{section}

\newcommand {\be} {\begin {equation}}
\newcommand {\ee} {\end {equation}}

\newcommand {\bes} {\begin {equation*}}
\newcommand {\ees} {\end {equation*}}

\newcommand{\es}[2] {\begin{equation} \label{#1} \begin{split} #2 \end{split} \end{equation}}

\newcommand{\beq}{\begin{equation}}
\newcommand{\eeq}{\end{equation}}

\def\be{ \begin{equation} }
\def\ee{ \end{equation} }

\begin{document}

\institution{PU}{Department of Physics, Princeton University, Princeton, NJ 08544}
\authors{Jeongseog Lee, Lauren McGough and Benjamin R.~Safdi
}

\title{R\'enyi Entropy and Geometry}

\abstract{ 
 Entanglement entropy in even dimensional conformal field theories (CFTs) contains well--known universal terms arising from the conformal anomaly. 
R\'enyi entropies are natural generalizations of the entanglement entropy that are much less understood. 
 Above two spacetime dimensions, the universal terms in the R\'enyi entropies are unknown for general entangling geometries.  We conjecture a new
 structure in the dependence of the four--dimensional R\'enyi entropies on the intrinsic and extrinsic geometry of the entangling surface.
 We provide evidence for this conjecture by direct numerical computations in the free scalar and fermion field theories.  The computation involves relating the four--dimensional free massless R\'enyi entropies across cylindrical entangling surfaces to corresponding three--dimensional massive R\'enyi entropies across circular entangling surfaces.  
  Our numerical technique
also allows us to directly probe other interesting aspects of three-dimensional R\'enyi entropy, including the massless renormalized R\'enyi entropy and calculable contributions to the perimeter law.
}

\maketitle

\section{Introduction}

Quantum entanglement is a powerful tool for characterizing the ground states of many-body quantum systems and continuum quantum field theories (QFTs).  A useful way of quantifying quantum entanglement is through the R\'enyi entropies~\cite{renyi0,renyi1} 
\es{Renyi}{
S^q ={ \log \tr \rho_R^q \over 1 - q} \,, \qquad q > 0 \,.
}
The reduced density matrix $\rho_R$ is computed by taking the trace of the ground state density matrix $\rho = | 0 \rangle \langle 0 |$ over the degrees of freedom living outside of the entangling region. 

A particularly interesting special case is the entanglement entropy (EE) (for reviews, see~\cite{cardyCFT1,Eisert:2008ur,Nishioka:2009un}):  
\es{EE}{
S  = \lim_{q \to 1} S^q = - \tr (\rho_R \log \rho_R) \,.
}  
At the conformal fixed points of even $d$-dimensional relativistic QFTs, the EE $S_\Sigma$ across any smooth $(d-2)$-dimensional entangling surface $\Sigma$ has a $\log \epsilon$ divergence, where $\epsilon$ is the short-distance cutoff, in addition to the power-like divergent terms $\propto \epsilon^{-2n}$, $n =1, \dots ,(d-2)/2 $.  When $d = 2$, the coefficient of the $\log \epsilon$ term is proportional to the central charge of the conformal field theory (CFT)~\cite{cardy0,cardyCFT1,cardyCFT2}.  In $d = 4$, Solodukhin~\cite{Solodukhin:2008dh} has proposed that\footnote{$A_\Sigma$ is the area of the entangling surface, and $E_2$ is the 2-dimensional Euler density on $\Sigma$.  The extrinsic curvature is given by $k^i_{ab}$, where $a,b = 1,2$ are local indices on $\Sigma$ and $i = 1,2$ label the two independent normal vectors.  The quantity $C^{ab}\,_{ab}$ is the trace of the pullback of the spacetime Weyl tensor onto $\Sigma$.} 
\es{Sd4}{
S_\Sigma = \alpha {A_\Sigma \over \epsilon^2} + \left[ {a \over 180} \int_\Sigma E_2 + {c \over 240 \pi} \int_\Sigma \left( \big( \tr k^2 - {1 \over 2} k^2 \big) - C^{ab}\,_{ab} \right) \right] \log \epsilon + O(\epsilon^0) \,,
}
where  the coefficients $a$ and $c$ are the two Weyl anomaly coefficients in 4-d CFT, normalized so that they both equal unity for a real scalar field.  There exists a similar proposal for the $\log \epsilon$ term in the EE in 6-d CFT~\cite{Safdi:2012sn}.  The proposal~\eqref{Sd4} has passed many non-trivial checks~\cite{Huerta:2011qi,Fursaev:2013fta,Bhattacharyya:2013gra,Miao:2013nfa,Dong:2013qoa,Erdmenger:2014tba}.

The R\'enyi entropies should also possess an expansion analogous to~\eqref{Sd4}.  In $(1+1)$-dimensional CFT, it is known that the universal term scales $\propto (1 + q^{-1})$ with the R\'enyi parameter $q$~\cite{cardy0,cardyCFT1,cardyCFT2}.  In $(3+1)$-dimensional CFT, we do not expect the universal term in the R\'enyi    entropy to be completely determined by $a$ and $c$~\cite{Hung:2011nu}.  However, in this note we provide evidence for non-trivial structure in that term.  As in~\eqref{Sd4}, the universal term in the R\'enyi entropy has an expansion in terms of geometric invariants on the entangling surface.   
The extrinsic curvature term in the expansion is notoriously hard to probe.  We conjecture a simple relation between the theory-dependent coefficient of this term and that of the Weyl tensor term that holds for all $q$ and for all CFTs. 

The rest of this note is organized as follows.  In Sec.~\ref{sec: univ} we present our conjectured relations in $(3+1)$-dimensional CFT R\'enyi entropy.  In Sec.~\ref{sec: num} we provide evidence for this conjecture through numerical computations of R\'enyi entropy with free fermion and scalar fields.  Also in that section, we numerically compute the CFT R\'enyi entropy across an $S^1$ entangling surface in the $(2+1)$-dimensional free scalar and fermion theories.  We show that the numerical calculations are consistent with the results obtained by conformally mapping the CFT R\'enyi entropy to the thermal free energy on $\mathbb{H}^2$~\cite{Klebanov:2011uf}.  In Sec.~\ref{Sec: D} we provide numerical evidence for a calculable, cutoff-independent contribution to the area-law term in $(2+1)$-dimensional massive R\'enyi entropy~\cite{Lewkowycz:2012qr}.  Details of the numerical calculations may be found in Appendices~\ref{sec: details} and~\ref{sec: expand}.

\section{Universal structure in R\'enyi entropy} \label{sec: univ}
In $(3+1)$-dimensional CFT, the $\log \epsilon$ term in the R\'enyi entropy should take the form~\cite{Fursaev:2012mp} 
\es{Rd4}{
\left. S_\Sigma^q \right|_{\log \epsilon} =  \left[ {f_a(q) \over 180} \int_\Sigma E_2 +{ f_b(q) \over 240 \pi} \int_\Sigma  \big( \tr k^2 - {1 \over 2} k^2 \big) -{ f_c(q) \over 240 \pi} \int_\Sigma C^{ab}\,_{ab} \right] \log \epsilon  \,,
}
where the functions $f_{a,b,c}(q)$ are independent of the entangling surface.  These functions must also approach the correct anomaly coefficients at $q = 1$ so as to reproduce~\eqref{Sd4}: $f_a(1)~=~a$, $f_{b,c}(1) = c$.\footnote{In addition, $f_a'(1) = -( 9 \pi^4 C_T) / 2$, where $C_T$ is the coefficient of the vacuum two-point function of stress-energy tensors, in units where $C_T = 1/(3 \pi^4)$ for a real scalar field~\cite{Perlmutter:2013gua}.} In the theory consisting of $n_0$ free real scalars and $n_{1/2}$ free Weyl fermions, for example, the functions $f_a(q)$ and $f_c(q)$ have been computed explicitly, yielding the results~\cite{Casini:2010kt,Fursaev:1993hm,DeNardo:1996kp,Fursaev:2012mp,Klebanov:2011uf} 
\es{fac}{
f_a(q) &= n_0 { (1+q)(1 + q^2) \over 4 q^3} + n_{1/2} { (1+q)(7 + 37 q^2) \over 16 q^3}   \,,  \\
f_c(q) &=  n_0 { (1+q)(1 + q^2) \over 4 q^3} + n_{1/2} { (1+q)(7 + 17 q^2) \over 16 q^3}   \,.
}  

Previous to this work, the function $f_b(q)$ has remained completely unknown for any theory away from $q = 1$.  In Sec.~\ref{sec: num} we compute the function $f_b(q)$ numerically for the free scalar and free fermion theories.  To within the numerical precision of our calculation, we find that $f_b(q) = f_c(q)$ in these examples.  This leads us to the following conjecture,
\begin{equation}
\text{\textbf{Conjecture:} }f_b(q) = f_c(q)\text{ in all $(3+1)$-dimensional CFTs.} 
\end{equation} 

A simple procedure exists for calculating entanglement entropy holographically in any dimension and for any entangling surface~\cite{Ryu:2006ef,rt,Klebanov:2007ws,Lewkowycz:2013nqa}.  
However, the holographic procedure for calculating R\'enyi entropy away from $q = 1$ is unknown for general geometries.  
It would be an interesting challenge to check our conjecture that $f_b(q) = f_c(q)$ holographically.

For free massless scalars and fermion fields, the universal term in the $(3+1)$-dimensional R\'enyi entropy across a cylinder of radius $R$ is related to the $1/(mR)$ term in the large $(mR)$ expansion of the $(2+1)$-dimensional R\'enyi entropy across a circle of radius $R$ for the corresponding massive theories.  A special case of this result was pointed out for the EE $(q = 1)$, and Huerta~\cite{Huerta:2011qi} made use of this result to provide numerical checks of~\eqref{Sd4}.  

We generalize the calculations in~\cite{Huerta:2011qi} away from $q = 1$.  As is the case with EE~\cite{ch1,Klebanov:2012yf,Huerta:2011qi,Casini:2005zv,Grover:2011fa}, it is natural to expect the R\'enyi entropies to have an expansion at large $(m R)$
\es{Sqm}{
S^q = \alpha^q {R \over \epsilon} + \beta^q \,m\, (2 \pi R) - \gamma^q + \sum_{n = 0}^\infty {C^q_{-1 - 2n} \over (mR)^{2n+1}} \,.
} 
The $\gamma^q$ are expected to be related to topological EE~\cite{preskill,levin}.  We discuss the $\beta^q$ in Sec.~\ref{Sec: D}.

By dimensional reduction, the coefficient $C^q_{-1 - 2n}$ in the free massive theories is related to the universal term in the R\'enyi entropy in the corresponding massless theory in $(2n + 4)$-dimensions.  The entangling surface in the higher-dimensional theory should be thought of as $T^{2n+1} \times S^1$, where $T^{2n+1}$ is the $(2n+1)$-torus, whose volume is taken to be large compared to $R^{2n+1}$, with $R$ the radius of the $S^1$.  This has been noted previously for the EE~\cite{Huerta:2011qi,Klebanov:2012yf,Safdi:2012sn}, and the arguments leading to the conclusion for the R\'enyi entropy are the same.  In particular, 
\es{c1}{
\left( C^q_{-1}\right)^\text{scalar} =- { \pi \over 240} f_b(q) \,, \qquad \left( C^q_{-1}\right)^\text{fermion} =- { \pi \over 480} f_b(q) \,.
} 
In deriving these relations, we used the fact that when the 4-d entangling geometry is $S^1 \times S^1$, where the first $S^1$ has a length $L$ much larger than the radius $R$ of the second $S^1$, then~\eqref{Rd4} evaluates to $S^q|_{\log \epsilon} = { f_b(q) \over 240} {L \over R} \log \epsilon$.  The fermion result has an extra factor of $1/2$ due to the spin multiplicity~\cite{Huerta:2011qi,Klebanov:2012yf,Safdi:2012sn}.

\section{Numerical  R\'enyi entropy} \label{sec: num}

By computing $C^q_{-1}$, we may use~\eqref{c1} to infer $f_b(q)$.  We compute $C^q_{-1}$ using the numerical method for calculating R\'enyi entropy proposed in~\cite{ch1}, which is a straightforward generalization of the Srednicki procedure for numerically calculating EE~\cite{Srednicki:1993im,2003JPhA...36L.205P,ch1,Huerta:2011qi}.  The Srednicki procedure has been used recently to numerically calculate EE with circular entangling surfaces in free massive theories~\cite{Huerta:2011qi,Liu:2012eea,Safdi:2012sn,Klebanov:2012va}.   

In the scalar theory, the method works by expanding the $(2+1)$-dimensional free massive field in angular momentum modes parameterized by the integer $n$.  We discretize the radial direction, and for each $n$ we compute the discrete Hamiltonian $H_n$.  Using $H_n$, we compute the contribution $S_n^q$ of the $n^\text{th}$ mode to the R\'enyi entropy: 
\es{SqExp}{ 
S^q&= S^q_0 + 2\sum\limits_{n=1}^\infty S^q_n \,.
}
This prescription works for integer and non-integer $q$, including $q < 1$.  The fermion computation is analogous, expect that $n$ is half-integer.  More details of this method are given in Appendix~\ref{sec: details}.

A key point is that the $S_n^q$ have the following expansions at large $n$, which are given in more detail in Appendix~\ref{sec: expand}:
\es{snasymp}{
\begin{array}{cc}
\text{Scalar:\,\,\, }
S_n^q \sim~
\begin{cases}
{1 \over n^{4q}} & q<1 \\
{1 \over n^4} & q\geq 1 
\end{cases} \,, & 
\qquad \, \, \text{Fermion:\,\,\, }
S_n^q \sim~
\begin{cases}
{1 \over n^{2q}} & q<1 \\
{1 \over n^2} & q\geq 1\,.
\end{cases}
\end{array}
}
These expressions demonstrate that our numerical 
methods break down when $q \leq \frac{1}{4}$ and $q \leq \frac{1}{2}$ for the scalar 
and the fermion theories, respectively, since for lower values of $q$ the sums over $n$ are not convergent.  We restrict our discussion to $q$ greater than these critical values.  

\begin{figure*}[tb]   
\leavevmode
\begin{center}$
\begin{array}{cc}
\scalebox{.62}{\includegraphics{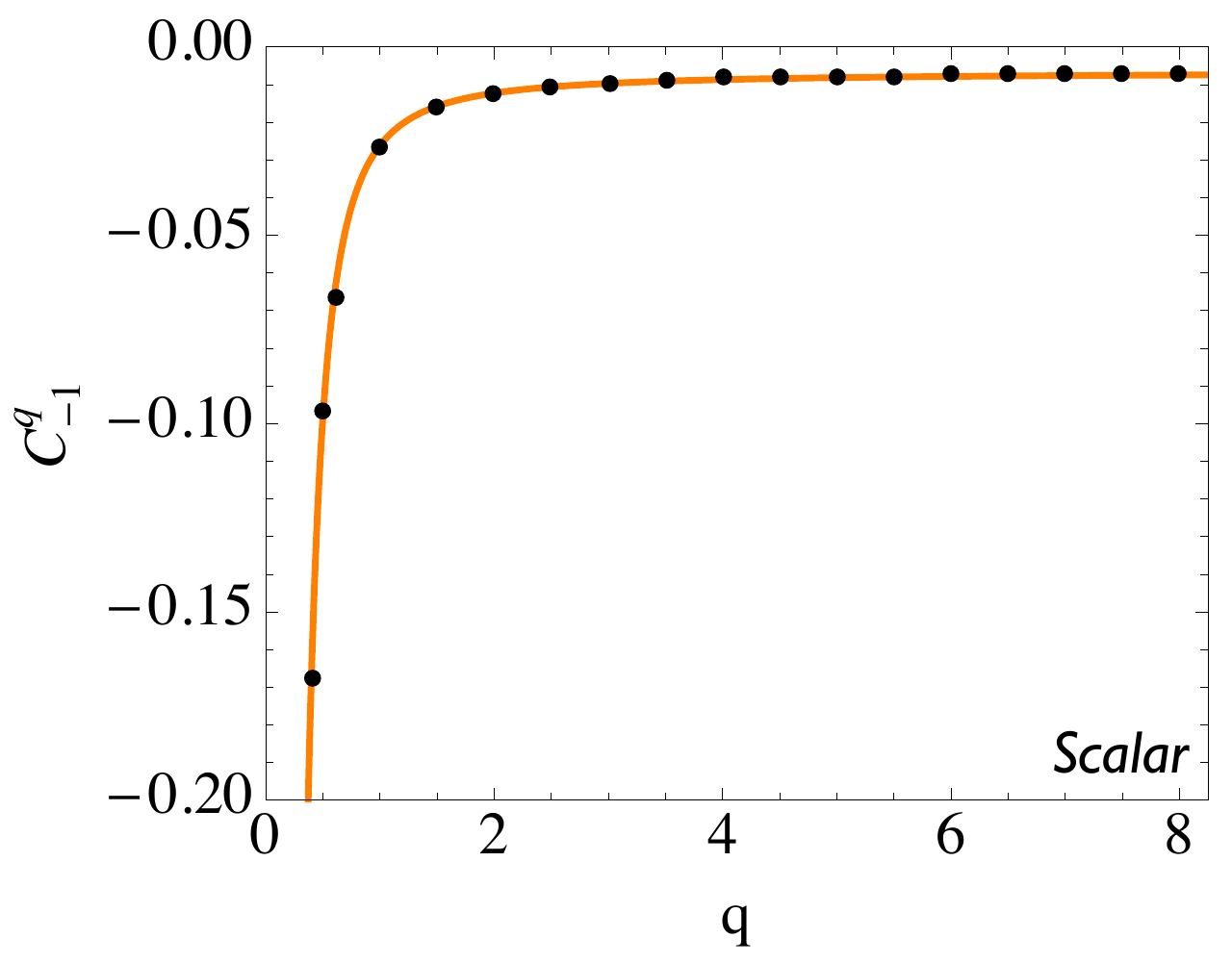}} & \scalebox{.62}{\includegraphics{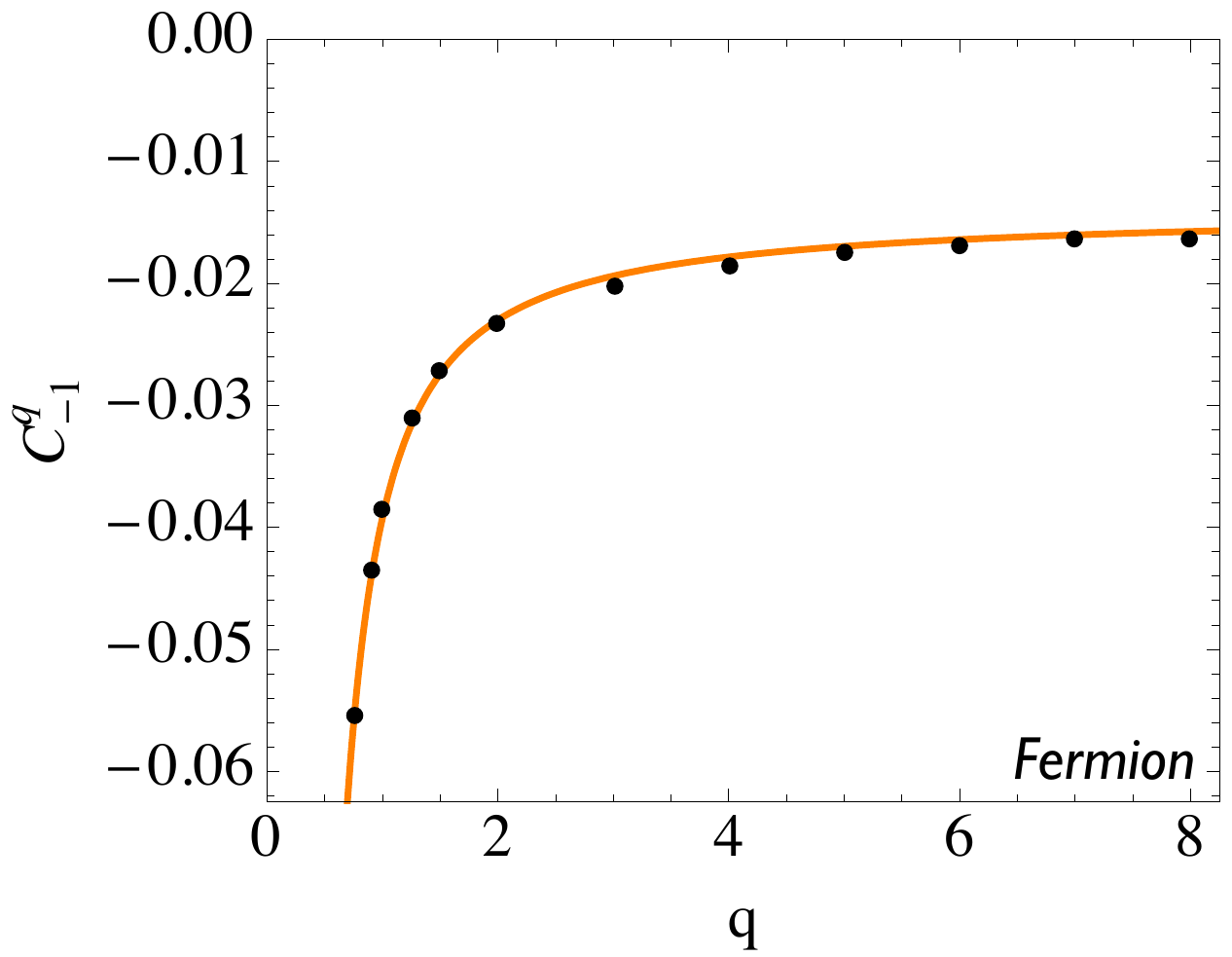}}   \\
 \end{array}$
\end{center}
\vspace{-.50cm}
\caption{
The coefficients $C^q_{-1}$ of the $1/(mR)$ term in the large $mR$ expansion of the R\'enyi entropy (see~\eqref{Sqm}) for the complex scalar (left) and Dirac fermion (right) theories.  These coefficients are related to the $f_b(q)$ coefficients appearing in~\eqref{Rd4} in the $(3+1)$-dimensional R\'enyi entropy for the massless CFTs through~\eqref{c1}.  The orange curves show the predictions from our conjecture that $f_b(q) = f_c(q)$, with the $f_c(q)$ given in~\eqref{fac}.  The black points are the results of the numerical calculations.  The numerical results agree with the analytic prediction to within  
3\% for the scalar and fermion theories for all $q$.
}
\vspace{-0.15in}
\label{fig: C}
\end{figure*}

After calculating the R\'enyi entropy $S^q(mR)$ as a function of the dimensionless parameter $mR$, we calculate a renormalized quantity 
\es{renormFq}{
{\cal F}^q (mR) = -S^q (mR)+ R\,\partial_R{S^q} (mR) \,.
}
The renormalized R\'enyi entropy has the nice features of being cutoff independent and of approaching the renormalized EE~\cite{Liu:2012eea,Casini:2012ei} as $q \to 1$.  We extract the coefficient $C^q_{-1}$ from the $1/(mR)$ term in the large-$mR$ expansion of this function.  
More specifically, for the scalar(fermion) we compute ${\cal F}^q (mR) $ over a range of $mR$ values between $0.7$ and $2.5$ ($2$ and $4$), and we fit this data to a function of the form 
\es{FitCm1}{
{\cal F}^q (mR)  \sim { \tilde C_{-1}^q \over m R} + {\tilde C_{-3}^q \over (m R)^3}  \,.
}
We take the $\tilde C_{-1}^q$ as our approximations to the $C_{-1}^q$.
The results of these computations are shown in Fig.~\ref{fig: C}.    
We find that the numerical calculations of $C^q_{-1}$ agree with the analytic predictions to within 
3\% across a broad range of $q$ in the free theories. 

We also extract the massless renormalized R\'enyi entropies, $S^q \equiv {\cal F}^q ( 0)$.  These quantities are of interest because the massless, free theories are conformal, 
so each $S^q$ may be computed analytically by mapping the computation of the R\'enyi entropy across the circle to the calculation of the thermal partition function on $ \mathbb{H}^2$~\cite{Hung:2011nu,headrick,Klebanov:2011uf}.  Taking the radius of $ \mathbb{H}^2$ to be $R$, the temperature is $1/(2 \pi R q)$.  The R\'enyi entropy $S^q$ is then related to the thermal free energy $F^q_\text{therm} = - \log |Z_q|$, where $Z_q$ is the Euclidean partition function on $S^1 \times \mathbb{H}^2$ (the $S^1$ has circumference $2 \pi R q$), through the relation~\cite{Casini:2010kt,Hung:2011nu}
\es{SandF}{
S^q = {q { F}^1_\text{therm} - { F}^q_\text{therm} \over 1 - q} \,.
}   
\begin{figure*}[tb]
\leavevmode
\begin{center}$
\begin{array}{cc}
\scalebox{.62}{\includegraphics{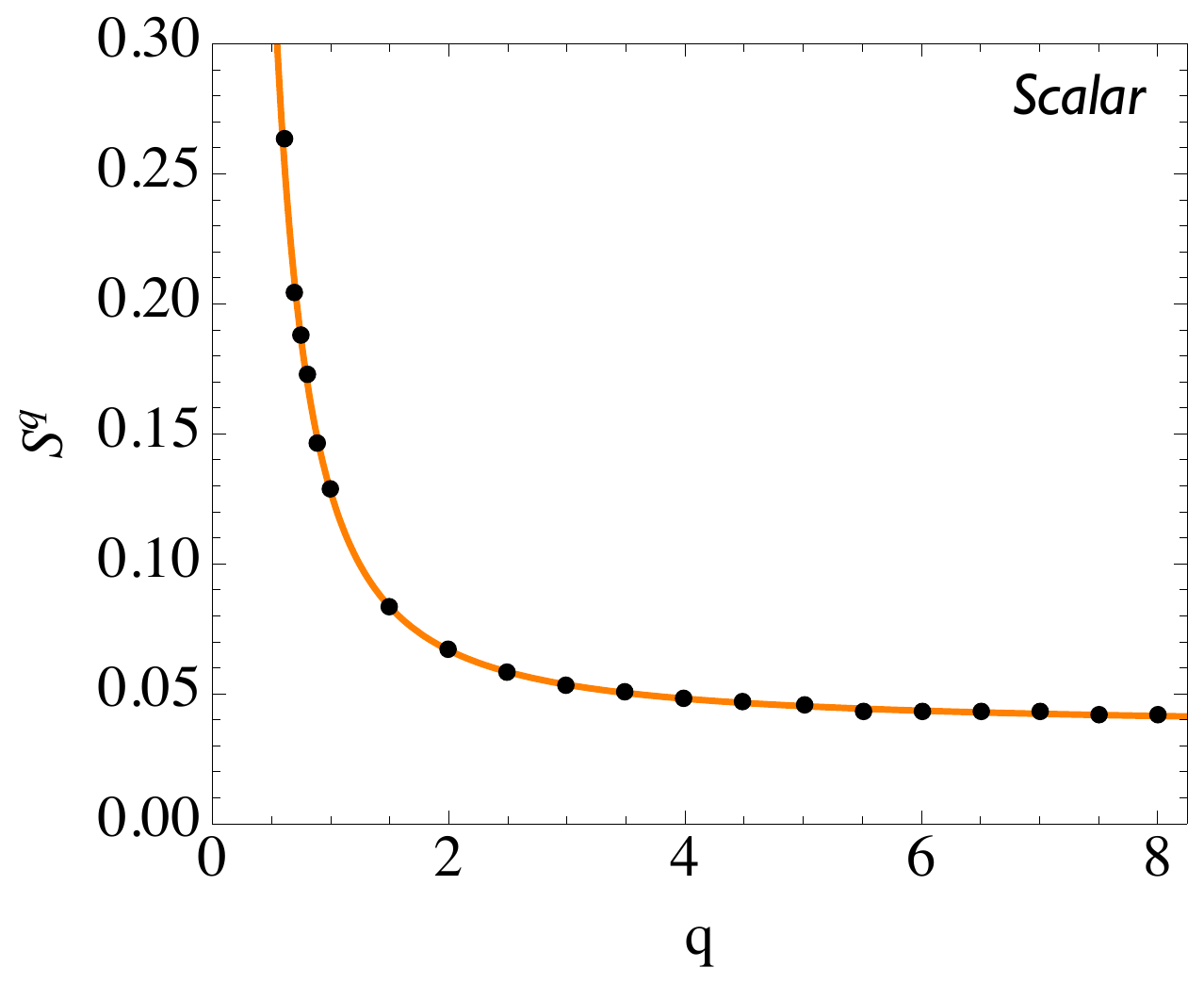}} & \scalebox{.62}{\includegraphics{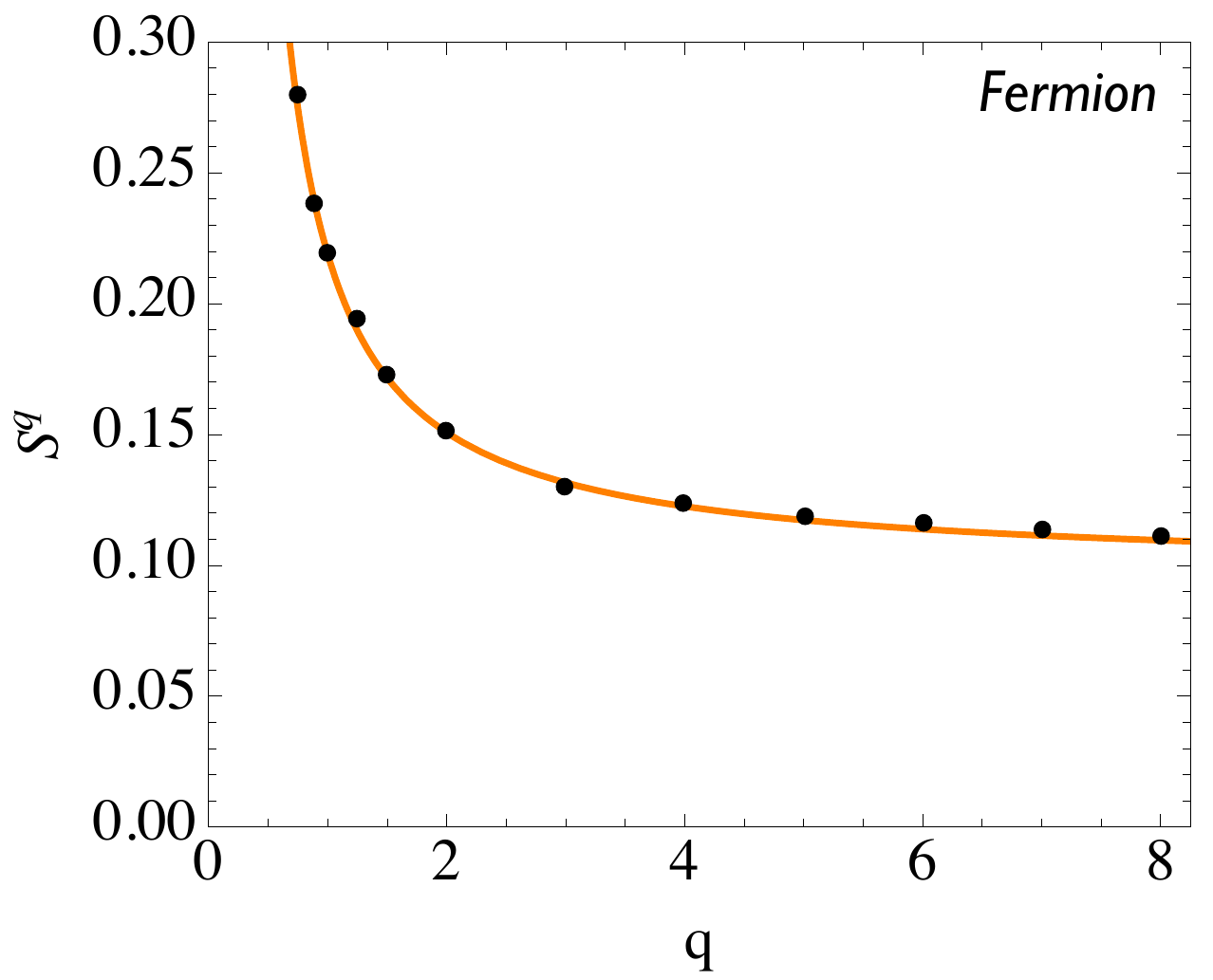}}  \\
 \end{array}$
\end{center}
\vspace{-.50cm}
\caption{The massless R\'enyi entropies $S^q$ in the free complex scalar (left) and Dirac fermion (right) theories as functions of the R\'enyi parameter $q$.  The orange curves are the analytic predictions coming from the mapping to $S^1 \times \mathbb{H}^2$ (see~\eqref{RenyiResult}).  The black points are the results of the numerical computation.  We find that the numerical results agree with the analytic predictions to within 2\% for the scalar and fermion theories across all $q$.
}
\vspace{-0.15in}
\label{fig: F}
\end{figure*}
 
 Recently, computations~\cite{Klebanov:2011uf} on $S^1 \times \mathbb{H}^2$  lead to the conjecture    
\es{RenyiResult}{
\big(F^q_\text{therm}\big)^\text{scalar} &= - \int_0^\infty d\lambda \, \textrm{tanh}(\pi\sqrt{\lambda})\textrm{log}(1-e^{-2\pi q \sqrt{\lambda}}) + q{3\zeta(3) \over 4\pi^2} \\
\big(F^q_\text{therm}\big)^\text{Dirac} &= 2 \int_0^\infty d\lambda \, \lambda \, \textrm{coth}(\pi\lambda)\textrm{log}(1+e^{-2\pi q \lambda}) + q{\zeta(3) \over \pi^2} 
}
for the complex scalars and Dirac fermions, respectively.  We stress that these computations involved various nontrivial elements, such as the regularization of the volume of $\mathbb{H}^2$ and the regularization of the sum over eigenvalues.    We test these results by direct numerical computations of $S^q$ at different values of $q$.
  In Fig.~\ref{fig: F} we compare the analytic predictions~\eqref{RenyiResult} to the numerical results.  Excellent agreement is seen, to within the numerical accuracy of the computation, over a wide range of $q$ for both the scalar and fermion theories.    
  
  \section{Calculable contributions to the perimeter law} \label{Sec: D}
  
  In the previous section, we numerically calculated the renormalized R\'enyi entropies ${\cal F}^q(mR)$ across a circle of radius $R$ in the free massive $(2+1)$-dimensional theories.  We extracted the coefficients $C^q_{-1}$ and $S^q$ from the ${\cal F}^q(mR)$ and then compared the results with analytic predictions.    
  However, there are interesting, physical aspects of the massive R\'enyi entropies $S^q(mR)$ that are not captured by the ${\cal F}^q(mR)$.  
  
  By construction, the ${\cal F}^q(mR)$ are not sensitive to terms in the $S^q(mR)$ that are linear in $R$; in particular, we cannot extract the $\beta^q$ coefficients (see~\eqref{Sqm}) from the renormalized R\'enyi entropies.  In this section, we extract the $\beta^q$ coefficients directly from the numerically-calculated R\'enyi entropies.  
  
   In~\cite{Hertzberg:2010uv} it was conjectured, using the results in~\cite{Solodukhin:2011gn,Fursaev:1994in,Fursaev:1995ef}, that $\beta^1 = - 1/12$ both for the massive real scalar and Dirac fermion theories.  That work assumed that for a general entangling surface $\Sigma$, the $\beta^1$-term in the entanglement entropy obeys a perimeter law,
  \es{SqmSig}{
S_\Sigma \supset \beta^1 \,m\, \ell_\Sigma \,,
}   
where $\ell_\Sigma$ is the perimeter of the entangling surface $\Sigma$.  To calculate $\beta^1$, the authors chose a simple entangling geometry -- the waveguide geometry -- where the entanglement entropy could be calculated explicitly using the heat kernel method.  
  
  Ref.~\cite{Huerta:2011qi} checked the analytic prediction for $\beta^1$ by numerically calculating the massive entanglement entropy in the free scalar and fermion theories.  Importantly,~\cite{Huerta:2011qi} used a circular entangling surface in flat spacetime. Perfect agreement was found between the numerical results and the analytic prediction.  This calculation provided evidence for the perimeter-law scaling~\eqref{SqmSig}.  
  
  It is natural to ask if the $\beta^q$ term in the R\'enyi entropy should also obey a perimeter law; if it does, then we may use the waveguide geometry to calculate the $\beta^q$ explicitly.  The results may then be applied to the circular geometry. The calculation in the waveguide geometry was performed in~\cite{Lewkowycz:2012qr}, and it was found that 
  \es{betaqFS}{
  \beta^q = -{1 + q\over 24 q} 
  }
  in both the real scalar and Dirac fermion theories. 
  
  Note that $\beta^q \propto (1 + q^{-1})$ has the same $q$ dependence as the universal term in $(1+1)$-dimension R\'enyi entropy~\cite{cardy0,cardyCFT1,cardyCFT2}.  This fact is not accidental, and it has a simple explanation.  Using the heat kernel method, the free-field R\'enyi entropy in the waveguide geometry is proportional to an appropriate integral over the heat kernel on $C_q \times S^1$, where $C_q$ is the two-dimensional cone with deficit angle $2 \pi (1 - q)$~\cite{Hertzberg:2010uv,Lewkowycz:2012qr}.  The heat kernel on this space factors into that on $C_q$ and that on $S^1$.  The two-dimensional R\'enyi entropy in the free theories may also be found by computing an appropriate integral over the heat kernel on $C_q$, and so it is not surprising that $\beta^q$ inherits the same $q$ dependence.     
  
  We check that~\eqref{betaqFS} applies also to circular entangling surfaces in flat spacetime by  generalizing the numerical calculation in~\cite{Huerta:2011qi} away from $q = 1$.  For each $q$ and each mass value $m$, we fit the numerically-computed R\'enyi entropy to the function 
  \es{fitBeta}{
  S^q =  b_1^q(m) R + b_0^q(m) + b_{-1}^q(m) {1 \over R}  \,,
  } 
  where $R$ is the radius of the circle.  The data $b_1^q(m)$ are dominated by the UV-divergent perimeter-law term, while the $\beta^q$ perimeter-law term makes a subleading contribution.  To separate the two contributions, we fit the data to the function
  \es{b1Fit}{
  b_1^q(m) = \tilde \alpha^q_2 \, m^2 + 2 \pi  \tilde \beta^q \, m + \tilde \alpha^q_0 
  \,.
  }
The coefficient $\tilde \alpha^q_{2}$ accounts for finite lattice-size corrections, and $\tilde \alpha^q_{0}$ is the cutoff-dependent contribution to the perimeter law.
    \begin{figure*}[tb]
\leavevmode
\begin{center}$
\begin{array}{cc}
\scalebox{0.9}{\includegraphics{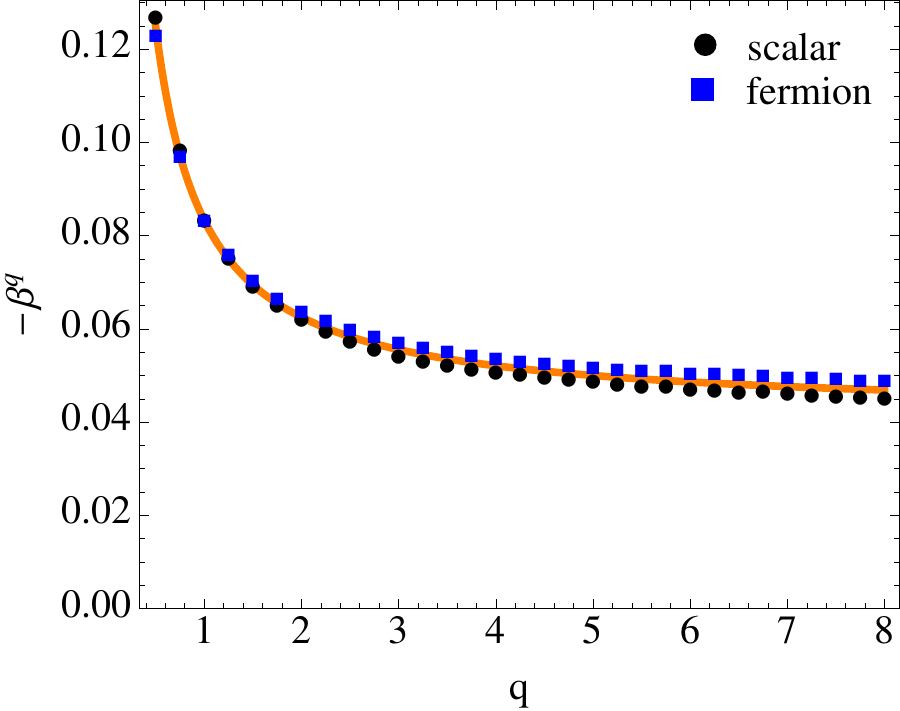}}   \\
 \end{array}$
\end{center}
\vspace{-.50cm}
\caption{
The coefficients $- \beta^q$ in the large-$mR$ expansion of the free-field R\'enyi entropy~\eqref{Sqm}.  An explicit computation of the $\beta^q$ in the wave-guide geometry~\cite{Lewkowycz:2012qr} combined with the assumption that the $\beta^q$-term obeys the perimeter law, as in~\eqref{SqmSig}, leads to the analytic prediction~\eqref{betaqFS} (solid orange) for both the real scalar and Dirac fermion theories.  Our numerical results for $\beta^1$ agree with~\eqref{betaqFS}  to better than $0.1$\%, while the $\beta^q$ at large $q$ agree with the analytic expression to within $\sim$2\%.
}
\vspace{-0.15in}
\label{fig: beta}
\end{figure*}

  To further improve the numerical precision, we repeat the calculation using lattices of varying size, between $N = 200$ and $N = 350$.  For each $q$, we fit the resulting data for $\tilde \beta^q$ to $\tilde \beta^q_0 + \tilde \beta^q_1 / N + \tilde \beta^q_2 / N^2$. We take $\tilde \beta^q_0$ as our approximation to $\beta^q$.  Our numerical results are compared to the prediction~\eqref{betaqFS} in Fig.~\ref{fig: beta}.  We find that $\beta^1$ deviates from $-1/12$ by less than $0.1$\% in both the scalar and fermion theories.  At large $q$ the deviation from~\eqref{betaqFS} is $\sim$2\%.
  We emphasize that the numerical results for $\beta^q$ are sensitive, at the few percent level, to the form of the fits~\eqref{fitBeta},~\eqref{b1Fit}, and the large-$N$ extrapolation.  In particular, in Fig.~\ref{fig: beta} it may be seen that at large $q$ the fermion results are systematically above the scalar ones. This gap decreases with increasing lattice size $N$.  While the large-$N$ extrapolation helps bring the two results into agreement, we are left with a small, residual error.  Practical limitations prevent us from further increasing the lattice size.
   
  The numerical procedures described in Sec.~\ref{sec: num} were subject to less difficulty. One reason for this is that the calculations in Sec.~\ref{sec: num} involved the renormalized entropies while the calculation of $\beta^q$ involves the R\'enyi entropies directly.  The process of separating the non-universal, cutoff-dominated component of the area law from the universal component is non-trivial and introduces additional lattice sensitivity.
  A direct, analytic calculation of the $\beta^q$ for the circular entangling surface
   would be useful.

\section*{Acknowledgments}
We thank A. Lewkowycz, S. Pufu and I.R. Klebanov for helpful discussions, and we thank I.R. Klebanov for suggesting the comparison in Fig.~\ref{fig: F} and for careful readings of this manuscript.   BRS and JL were supported in part by the US NSF grant PHY-1314198.  JL also received support from the Samsung Scholarship. LM was supported by a NSF GRFP under grant DGE-1148900.   
\appendix

\section{The numerical technique } \label{sec: details}

In this Appendix we give details of the Srednicki procedure for calculating R\'enyi entropy across a circular entailing surface in free 3-dimensional CFT~\cite{Srednicki:1993im,2003JPhA...36L.205P,ch1,Huerta:2011qi}.  We begin with an account of the procedure for the scalar theory. 
For our purposes, the Hamiltonian is most conveniently 
expressed by expanding in modes of 
integer angular momentum $n$ 
and discretizing the radial direction into $N$ units. 
The method relies on the observation that 
the resulting Hamiltonian takes the form 
\es{}{
H = \sum_n H_n \,, \qquad H_n= \frac{1}{2}\sum_i \pi_i^2 + \frac{1}{2}\sum_{ij} \phi_i K^{ij}_n \phi_j \, ,
}
where $\pi_i$ is the conjugate momentum to $\phi_i$, and 
$i, j$ run from $1,\ldots,N$. The matrix $K_n^{ij}$ has non-zero elements~\cite{Huerta:2011qi} 
\es{nZscalar}{
K_n^{11} = \frac{3}{2}+n^2+m^2 \,, \qquad 
K_n^{ii} = 2 + \frac{n^2}{i^2} + m^2 \,,\qquad
K_n^{i,i+1} = K_n^{i+1,i} = -\frac{i+1/2}{\sqrt{i(i+1)}}\, .
}

For each $n$, the two-point correlators $X_{ij} = \langle \phi_i \phi_j \rangle$ 
and $P_{ij} = \langle \pi_i \pi_j \rangle$ are directly related to $K$,
\es{XPscalar}{
X_n = \frac{1}{2} \left(K_n^{1/2}\right) \,, \qquad P_n = \frac{1}{2} \left(K_n^{-1/2}\right) \,,
}
and are thus easily computed.
We find the $q^\text{th}$ R\'enyi entropy across a circle of radius $R = r+\frac{1}{2}$ 
in lattice units by constructing the truncated matrices 
$X^r\equiv\big(X_{ij}\big)_{1\leq i,j \leq r}$ and 
$P^r\equiv \big(P_{ij}\big)_{1\leq i,j \leq r}$, for each $n$. The R\'enyi entropy is 
then given by~\eqref{SqExp} with
\es{Renyin}{
S^q_n &= {1 \over 1-q} \tr \log \left[ \left(\sqrt{X_n^r P_n^r} + {1 \over 2} \right)^q - \left(\sqrt{X_n^r P_n^r} - {1 \over 2}\right)^q\, \right] \,.
}

The fermion computation is similar. We expand the fermion field in half-integer angular momentum modes, labeled by $n$, and write the Hamiltonian as $H = \sum_n H_n$.  As before, the radial direction is discretized into $N$ units. 
For each $i=1,\ldots,N$, we decompose the Dirac spinor 
$\psi_i$ as $\psi_i = (u_i,v_i)^T$. Each $H_n$ takes the form 
\begin{align}
H_n &= \sum_{i,j=1}^N \begin{pmatrix}u_i^*&v_i^*\end{pmatrix} M^{i,j}_n \begin{pmatrix} u_j\\v_j\end{pmatrix},
\end{align}
where the $M^{ij}_n$ are known $(2\times 2)$ matrices with entries $\big(M^{ij}_n\big)_{\alpha \beta}$, $\alpha,\beta = 1,2$. 
It is convenient to reorganize the data for each $n$ into a single $(2N\times 2N)$ 
matrix $\tilde{M}^{2k+\alpha-2,2l+\beta-2}_n\equiv \big(M^{kl}_n\big)_{\alpha\beta}$.
The nonzero entries of the matrices $\tilde{M}_n$ are given by~\cite{Huerta:2011qi} 
\es{nonzeroF}{
\begin{array}{lll}
\tilde{M}^{kk}_n = (-1)^{k+1} m \,, \qquad & \tilde{M}^{1,2}_n = i\left(n+\frac{1}{2}\right) \,, \qquad & \tilde{M}^{2k-1,2k}_n = i \frac{n}{k} \,, \\
\tilde{M}_n^{2k-1,2k+2} = -\frac{i}{2} \,, \qquad & \tilde{M}_n^{2k-1,2k-2} = \frac{i}{2} \,, \qquad & \tilde{M}^{2,1}_n = -i\left(n+\frac{1}{2}\right) \,, \\
\tilde{M}^{2k,2k-1}_n = -i \frac{n}{k} \,, \qquad & \tilde{M}_n^{2k,2k-3} = \frac{i}{2}\,, \qquad & \tilde{M}_n^{2k,2k+1} = -\frac{i}{2} \,.
\end{array}
}

As in the scalar case, each $n$ is associated with a 
matrix of correlators $C_{ij} = \langle \psi_i \psi_j^{\dagger} \rangle$; 
in the fermion case, the $C_n$ are given by~\cite{Huerta:2011qi}
\begin{equation}
C_n = \Theta\left(-\tilde{M_n}\right).
\end{equation}
To compute the $q^\text{th}$ R\'enyi entropy across a circle of radius 
$R = r+\frac{1}{2}$, we define 
a truncated $C^r_n = \big(C_{ij}\big)^{1\leq i,j \leq r}_n$ for each $n$.
The R\'enyi entropy $S^q(R)$ is then given by 
\es{SqRferm}{
S^q(R) = \sum_n S^q_n \,, \qquad S^q_n = \frac{1}{1-q} \tr \log \left[ (1-C_n^r)^q + (C_n^r)^q \right] \,.
}

For both the scalar and fermion theories, we take our radial lattice to have $N=200$ points. 
For the scalar case, we study entangling circles of radii $30 \leq r \leq 50$ in lattice units, and 
for the fermion we take $20\leq r\leq80$. 
We study the massive scalar with mass $m=0.01 k$, $0\leq k\leq 8$, and the massive fermion with mass $m=0.005 k$, $0\leq k\leq 20$.
We compute the $S_n^q$ for $0\leq |n| \leq 2000$ for the scalar and $\frac{1}{2}\leq |n| \leq \frac{4001}{2}$ for the fermion. 

The primary sources of error in the numerical method
are finite lattice size effects and finite angular momentum cutoff effects.  We address the latter source of error by summing the asymptotic expansions of the $S_n$ at large $n$, given in Appendix~\ref{sec: expand}, from the angular momentum cutoff to infinity.
Finite lattice size effects are most pronounced for small angular momentum modes~\cite{Lohmayer:2009sq}. 
To adjust for this, for small $n$ we compute the $S^q_n$ on larger lattices and extrapolate to obtain the value 
as the lattice size approaches infinity. Specifically,
for the scalar we carry out the corrections for angular momentum modes $n=0,1  \dots ,5$ 
on lattices of size $N=200+10 \cdot i$ for $i=0, \dots, 49$; for the fermion, 
modes $n = \frac{1}{2}, \frac{3}{2},\ldots,\frac{15}{2}$ on lattices of size $N=200+10 \cdot i$, with $i=0, 1, \dots, 15$. 
 Denoting the lattice size as $N$, we fit the resulting data to
\es{snlattice}{
S_n = a + {b_1 \over N^2} + {b_2 \log N \over N^2} + {c_1 \over N^4} + {c_2 \log N \over N^4} + {d_1 \over N^6} + {d_2 \log N \over N^6} 
}
and extrapolate to $N \rightarrow \infty$ to obtain the lattice-size-corrected value.

\section{The $S_n$ at large $n$} \label{sec: expand}

It was shown in~\cite{Klebanov:2012va} how, in the scalar theory, one may determine the leading, relevant eigenvalues of $\sqrt{X_n^r P_n^r}$ that go into determining $S^q(R)$ (see~\eqref{Renyin}) in a $1/n$ expansion at large $n$.  A key point in the derivation is that the matrices $K_n$ are diagonal to leading order in $1/n$.  The result is that the matrix $\sqrt{X_n^r P_n^r} + {1\over2}$ has one eigenvalue equal to~\cite{Klebanov:2012va} 
\es{eigenS1}{
1 + {r^2 (r+1)^2 \over 16 n^4} + O(1/n^6) \,,
}
with all other eigenvalues equal to unity to higher order in $1/n$.  Similarly, the leading eigenvalue (away from zero) of the matrix $\sqrt{X_n^r P_n^r} - {1\over2}$ is~\cite{Klebanov:2012va}
\es{eigenS2}{
{r^2 (r+1)^2 \over 16 n^4} + O(1/n^6) \,.
}
Notice that the leading-order terms in the eigenvalues are independent of the mass $m$.

We may calculate the $1/n$ expansion of the $S^q_n$ by substituting the eigenvalues~\eqref{eigenS1} and~\eqref{eigenS2} into~\eqref{Renyin}.  The leading-order behavior of the expansion depends on whether or not $q >1$.  If $q > 1$, we find
\es{qB1S}{
S_n^q = {q \over 1 - q} {r^2(r+1)^2 \over 16 n^4} + o(1/n^4) \,,
}
while if $q  < 1$,
\es{qL1S}{
S_n^q = - {1 \over 1- q} \left( {r^2 (r+1)^2 \over 16 n^4} \right)^q + o(1/n^{4q}) \,.
}
 Note that when $q = 1$ there is $(\log n)/ n^4$ term in the expansion, while this term is absent away from $q = 1$.
 
The fermion computation is similar.  The matrices $\tilde M_n$ are block diagonal to leading order in $1/n$.  The $2 \times 2$ blocks are indexed by $k = 1, \dots, N$.  To leading order in $1/n$, the $k^\text{th}$ block 
\es{}{
\left( \begin{array}{cc}
0 & {i n \over k} \\
- {i n \over k} & 0 
\end{array}
\right) 
}
is diagonalized by $V^{(1)}_k = ( i, 1)$, with eigenvalue ${n \over k}$, and $V^{(2)}_k = (- i, 1)$, with eigenvalue $-{n \over k}$.  Let $D = \text{Diag} \left(n, - n, {n \over 2}, - {n \over 2}, \dots, {n \over N}, - {n \over N} \right)$ be the diagonal matrix of the eigenvalues of $\tilde M_n$.  To construct the matrix $C_n$, we should evaluate $\Theta(- D) = \text{Diag} (0, 1, \cdots, 0, 1 )$.  At large $n$, this result does not receive $1/n$ corrections.  From this discussion, it is straightforward to see that the eigenvalues of the reduced matrix $C_n^r$, with $r$ even, are equal to either $0$ or $1$ to leading order in $1/n$.

The eigenvectors may be corrected order by order in $1/n$.  To next-to-leading order, we find that 
\es{}{
V^{(1)}_k &= \left( 0, \dots,0 , - i {\alpha_{k-1} \over 2 n}, {\alpha_{k-1} \over 2n} , i ,1 - {k m \over n}, {i \alpha_k \over 2n}, -{ \alpha_k \over 2n}, 0, \dots, 0\right) \,,  \\
V^{(2)}_k &= \left( 0, \dots,0 , + i {\alpha_{k-1} \over 2 n}, {\alpha_{k-1} \over 2n} ,- i ,1 + {k m \over n}, -{i \alpha_k \over 2n}, -{ \alpha_k \over 2n}, 0, \dots, 0\right) \,,
}
with $\alpha_k = k(k+1) / (2k+1)$ and the first non-zero entries above sitting $2k-3$ positions to the right.  Only the eigenvectors $V^{(1)}_{r/2}$ and $V^{(2)}_{r/2}$ are relevant for determining the two leading eigenvalues of the the matrix $C_n^r$.  A straightforward calculation shows that these eigenvalues are 
\es{twoFerm}{
1 - {r^2 (r+2)^2 \over 64 (r+1)^2} {1 \over n^2} \,, \qquad {r^2 (r+2)^2 \over 64 (r+1)^2} {1 \over n^2} \,.
}
All other eigenvalues are equal to either $0$ or $1$ to higher order in $1/n$.  Moreover, the eigenvalues are independent of $m$ at order $1/n^2$.  The $m$-dependence arises at higher order.  Substituting these eigenvalues into~\eqref{SqRferm}, we may construct the large $n$ expansion of the $S_n^q$.  When $q > 1$, we find
\es{qBqF}{
S_n^q = - { q \over 1 - q} {r^2 (r+2)^2 \over 32 (r+1)^2} {1 \over n^2} + o(1/n^2) \,,
}
while when $q < 1$,
\es{qLqF}{
S_n^q = { 2 \over 1 - q} \left( {r^2 (r+2)^2 \over 64 (r+1)^2} {1 \over n^2} \right)^q + o(1/n^{2q}) \,.
}

\bibliographystyle{ssg}
\vspace{-.39in}
\bibliography{draft}

\end{document}